# Electrically tunable spin-orbit interaction in an InAs nanosheet

Furong Fan[1], Yuanjie Chen[1], Dong Pan[2], Jianhua Zhao[2], and H. Q. Xu[1,3,*]

[1]*Beijing Key Laboratory of Quantum Devices, Key Laboratory for the Physics and Chemistry of Nanodevices, and School of Electronics, Peking University, Beijing 100871, China*

[2]*State Key Laboratory of Superlattices and Microstructures, Institute of Semiconductors, Chinese Academy of Sciences, P.O. Box 912, Beijing 100083, China*

[3]*Beijing Academy of Quantum Information Sciences, Beijing 100193, China*

*Corresponding author; email: hqxu@pku.edu.cn

(Date: April 30, 2022)

## ABSTRACT

We report on an experimental study of the spin-orbit interaction (SOI) in an epitaxially grown free-standing InAs nanosheet in a dual-gate field-effect device. Gate-transfer characteristic measurements show that independent tunings of the carrier density in the nanosheet and the potential difference across the nanosheet can be efficiently achieved with use of the dual gate. The quantum transport characteristics of the InAs nanosheet are investigated by magnetoconductance measurements at low temperatures. It is shown that the electron transport in the nanosheet can be tuned from the weak antilocalization to the weak localization and then back to the weak antilocalization regime with a voltage applied over the dual gate without a change in carrier density. The spin-orbit length extracted from the magnetoconductance measurements at a constant carrier density exhibits a peak value at which the SOI of the Rashba type is suppressed and the spin relaxation due to the presence of an SOI of the Dresselhaus type in the nanosheet can be revealed. Energy band diagram simulations have also been carried out for the device at the experimental conditions and the physical insights into the experimental observations have been discussed in light of the results of simulations.

Low-dimensional III-V narrow bandgap semiconductor nanostructures have attracted significant interests due to their potential applications in nanoelectronics,[1,2] infrared optoelectronics,[3,4] spintronics,[5-7] and quantum electronics.[8-13] Among them, strong spin-orbit interaction (SOI) and large Landé $g$-factor have made InAs and InSb nanostructures widely used in realization of novel devices such as spin transistors,[14,15] spin-orbit qubits,[16,17] and topological quantum devices.[18–21] The SOI originates from the nonrelativistic approach to the relativistic Dirac equation and describes that a moving (orbit) charged particle in an electric field experiences an effective magnetic field and thus the spin and momentum of the charged particle are coupled with the coupling strength being related to the electric field. In crystalline materials, SOIs appear dominantly due to bulk inversion asymmetry (known as the Dresselhaus type) and structural inversion asymmetry (known as the Rashba type).[22,23] The Rashba SOIs provide a possibility to manipulate spins in the materials by electrical means and thus have potential applications in building spin-based devices, such as Datta-Das spin transistors,[14,24] and fast controllable spin qubits.[25-27] Recently, epitaxial growth techniques of single-crystalline, III-V narrow bandgap, semiconductor nanostructures have been successfully extended to grow free-standing InAs nanosheets.[28] Preliminary study has experimentally demonstrated the existence of strong SOI in the nanosheets.[9] However, due to the employment of a single-gate device configuration, the study has failed to reveal the origin and an efficient control of the SOI in the InAs nanosheets.

In this work, we report on a transport measurement study of a dual-gate field-effect device made from an epitaxially grown free-standing InAs nanosheet and on demonstration of that the SOI presented in the InAs nanosheet is strongly tunable with use of the dual gate. We will show that the electron transport in the InAs nanosheet can be efficiently tuned from the weak antilocalization (WAL) to the weak localization (WL) and then back to the WAL regime without tuning carrier density and that the spin-orbit length extracted at a constant carrier density exhibits a peak value, at which the Rashba SOI is suppressed and thus the Dresselhaus SOI can be revealed. We will also present energy band diagram simulations for the layer structure of the device and discuss the physical insights of the experimental observations.

The free-standing, single-crystalline and pure-phase InAs nanosheet investigated in this work is grown via molecular beam epitaxy (MBE) on a Si (111) substrate. The growth is carried out under an indium-rich condition, at a temperature of 545 °C for 40 min, with the indium and arsenic fluxes fixed at $9.3 \times 10^{-7}$ and $5.9 \times 10^{-6}$ mbar, respectively (see Ref. 28 for further details about the nanosheet growth). Figure 1(a) shows a side-view scanning electron microscope (SEM) image of as-grown free-standing InAs nanosheets on the growth Si (111) substrate. The thicknesses of these InAs nanosheets are about 15–30 nm. Figure 1(b) is a high-resolution transmission electron microscope (HRTEM) image of an MBE-grown InAs nanosheet and the inset of Fig. 1(b) shows an electron diffraction (SAED) pattern taken at a selected area of the nanosheet. The HRTEM measurements reveal that these InAs nanosheets are of high-quality wurtzite crystals.

The device fabrication starts with transferring as-grown InAs nanosheets from the growth substrate onto a heavily n-doped Si substrate capped with a 200-nm-thick $SiO_2$ layer on top. The Si substrate (contacted by a gold film at the bottom) and the $SiO_2$ layer are later served as the bottom gate and the bottom-gate dielectric. We select a few InAs nanosheets on the $SiO_2$-capped Si substrate and locate them by SEM. On each selected InAs nanosheet, four contact electrodes are fabricated by pattern definition via electron-beam lithography (EBL), deposition of a Ti/Au (5/90 nm in thickness) metal bilayer via electron-beam evaporation (EBE) and lift-off. Here, we note that before metal deposition, the sample is briefly merged into a diluted $(NH_4)_2S_x$ solution in order to remove the native oxide layer on the nanosheet in the contact areas and to passivate the obtained fresh surfaces. Then, after another step of EBL, a 20-nm-thick $HfO_2$ dielectric layer covering the entire InAs nanosheet is fabricated by atomic layer deposition and lift-off. Finally, a Ti/Au (5/120 nm in thickness) metal bilayer top gate is fabricated again by EBL, EBE and lift-off. Figure 1(c) shows a top-view SEM image (in false color) of a fabricated dual-gate field-effect InAs nanosheet device (named as Device1) studied in this work and Fig. 1(d) displays a schematic for the layer structure of the device. In the device, the average width ($W$) of the InAs conduction channel is ~410 nm and the channel length ($L$), defined by the separation between the two inner contact electrodes, is ~830 nm. The thickness of the InAs nanosheet determined by atomic force microscopy measurements after the device fabrication is in a range of 17 to 21 nm. The fabricated device is characterized by gate transfer characteristic measurements and low-field magnetotransport measurements in a four-probe circuit setup, as shown in Fig. 1(c), using a standard lock-in technique, in which a 17-Hz, 10-nA excitation current ($I$) is supplied through the two outer contact electrodes and the voltage drop ($V$) between the two inner contact electrodes is recorded. The channel conductance ($G$) is obtained through $G = I/V$. All the measurements are performed in a physical property measurement system equipped with a uniaxial magnet at low temperatures $T$ and at magnetic fields $B$ applied perpendicular to the InAs nanosheet plane.

Figure 2 shows the measured InAs nanosheet channel conductance ($G$) for Device 1 [shown in Fig. 1(c)] as a function of bottom-gate voltage $V_{BG}$ and top-gate voltage $V_{TG}$ at a temperature of $T = 2$ K and zero magnetic field. It is seen that the dual-gate device exhibits a typical $n$-type transistor characteristic and operates in a depletion mode. It is also seen that the conductance $G$ can be tuned by both gates. However, the top gate exhibits a much strong capacitive coupling to the InAs conduction channel than the bottom gate, which is consistent with the fact that the top gate has a much shorter distance to the InAs nanosheet and a gate dielectric with a larger dielectric constant than the bottom gate. The pink, purple and orange solid lines in Fig. 2 represent the constant conductance contours of $G \sim 12$, $\sim 7$ and $\sim 4$ e$^2$/h, respectively. Moving along each constant conductance contour, the carrier density in the InAs nanosheet can be assumed to stay approximately at a constant value, but dual-gate voltage $V_D$ (defined as $V_D = V_{BG} - V_{TG}$) is continuously changed.

The carrier density ($n$), Fermi wavelength ($\lambda_F$), and mean free path ($L_e$) in the InAs nanosheet of a dual-gate device can be estimated from the gate transfer characteristic measurements (cf. Section I in Supplementary Information). The estimated values of $n$, $\lambda_F$, and $L_e$ for the InAs nanosheet in Device 1 at three different conductance values of $\sim 12$, $\sim 7$ and $\sim 4$e$^2$/h are given in Table S1. The estimated values of $\lambda_F \sim 22$–33 nm are comparable to or larger than the thickness of the InAs nanosheet, but much smaller than the lateral sizes of the nanosheet. Thus, the electron transport in the InAs nanosheet at these conductance values could be considered as of a two-dimensional (2D) nature. The estimated values of $L_e$ in the InAs nanosheet are also much smaller than the InAs nanosheet channel length $L$. The electron transport in the InAs nanosheet is thus in the diffusive regime. It is seen that a larger value of $L_e$ appears at a higher carrier density $n$ in the InAs nanosheet, which is consistent with the fact that a higher carrier density gives rise to a stronger screening of scattering centers for conduction electrons in the nanosheet.

In a 2D diffusive electron conducting system, the transport characteristics, such as phase coherence length $L_\varphi$, spin-orbit length $L_{SO}$, and mean free path $L_e$, can be extracted from the measurements and analyses of the low-field magnetoconductance. At low temperatures, the electron interference gives rise to a quantum correction to the classical conductance. Hikami, Larkin, and Nagaoka (HLN) showed that by taking the quantum correction into account, the magnetoconductance, $\Delta G = G(B) - G(B = 0)$, at low magnetic fields can then be expressed as,[29]

$$\Delta G(B) = -\frac{e^2}{\pi h}\left[\frac{1}{2}\Psi\left(\frac{B_\varphi}{B}+\frac{1}{2}\right) + \Psi\left(\frac{B_e}{B}+\frac{1}{2}\right) - \frac{3}{2}\Psi\left(\frac{(4/3)B_{SO}+B_\varphi}{B}+\frac{1}{2}\right) - \frac{1}{2}\ln\left(\frac{B_\varphi}{B}\right) - \ln\left(\frac{B_e}{B}\right) + \frac{3}{2}\ln\left(\frac{(4/3)B_{SO}+B_\varphi}{B}\right)\right], \qquad (1)$$

where $\Psi(x)$ is the digamma function, and $B_i$ (i = φ, SO, e) are the characteristic fields for different scattering mechanisms. The characteristic transport lengths are related to these characteristic fields by $B_i = \hbar/(4eL_i^2)$.

Figure 3(a) shows the measured low-field magnetoconductance $\Delta G$ of Device 1 at $T = 2$ K and at different dual-gate voltages $V_D$ taken along the conductance contour of $G \sim 12e^2/h$ (the pink line in Fig. 2). It is seen that at $V_D = -3.36$ V, $\Delta G$ displays a peak near zero magnetic field, i.e., a WAL characteristic,[30] implying the presence of a strong SOI in the InAs nanosheet. As $V_D$ decreases from –3.36 V to –6.28 V, the measured $\Delta G$ shows a transition from a peak-like structure to a dip-like structure near zero magnetic field, i.e., a transition from the WAL to a WL characteristic.[31] However, as $V_D$ is further decreased from –6.28 V to –7.62 V, the WL characteristic is gradually suppressed and $\Delta G$ turns to show the WAL characteristic again. This continuously tuning of the electron transport in the InAs nanosheet from the WAL to the WL and back to the WAL characteristic again is for the first time ever demonstrated experimentally in a planar semiconductor nanostructure.

To quantitatively analyze the tuning of the SOI in the InAs nanosheet, we fit the measured $\Delta G$ data to the HLN formula [Eq. (1)]. The solid lines in Fig. 3(a) are the results of the fits. Figure 3(b) displays the extracted values of $L_\varphi$, $L_{SO}$, and $L_e$ from the fits as a function of $V_D$ at the fixed conductance value of $G \sim 12e^2/h$. It is seen that the extracted $L_\varphi$ displays a weak dependence on $V_D$ and stays at $L_\varphi \sim 214$ nm. This is consistent with the fact that the electron dephasing in the nanosheet is dominantly originated from electron-electron interaction at low temperatures, which remains approximately unchanged without a change in carrier density. It is also seen that the extracted $L_e$ is also weakly dependent on $V_D$ and has a value of $L_e \sim 82$ nm. This value is consistent with the value of ~85 nm extracted from the gate transfer characteristic measurements. However, the extracted $L_{SO}$ exhibits a strong dependence on $V_D$. As $V_D$ decreases from –3.36 V to –6.28 V, $L_{SO}$ increases from ~83 nm to ~268 nm. As $V_D$ further decreases, $L_{SO}$ turns to decrease and has a value of ~165 nm at $V_D$ = –7.62 V. A smaller $L_{SO}$ indicates a stronger spin relaxation. Thus, the electrons in the InAs nanosheet of Device 1 exhibit a weak spin relaxation at $V_D$ = –6.28 V, but the spin relaxation gets stronger when $V_D$ moves away from the value of $V_D$ = –6.28 V. Similar low-field magnetoconductance $\Delta G$ measurements have also been carried out for Device 1 along the constant conductance contours of $G$ ~ 7$e^2$/h and $G$ ~ 4$e^2$/h at $T$ = 2 K (see Section IV and Fig. S4 in Supplementary Information). These measurements display again the transition from the WAL to the WL and then back to the WAL characteristic with decreasing dual-gate voltage $V_D$ along a constant conductance contour (i.e., a constant carrier density). The extracted spin-orbit length $L_{SO}$ at each constant carrier density again appears to be greatly tunable by dual-gate voltage $V_D$ and exhibits a peak value at which the strength of the SOI is at a minimum value.

Since, in all the above magnetotransport measurements, the tuning of $V_D$ is made along a constant conductance contour and thus approximately at a constant carrier density in the InAs nanosheet, a change in $L_{SO}$ with decreasing $V_D$ is most probably due to a change in the perpendicular electric field and thus a tuning of the Rashba SOI in the InAs nanosheet. To support this, we have simulated the energy band diagrams for the HfO$_2$/InAs/SiO$_2$ sandwiched layer structure in Device 1 at different dual-gate voltages and carrier densities based on Poisson's equation with the input material parameters of HfO$_2$, InAs, and SiO$_2$ as listed in Table S2 of Supplementary Information (see Section III in Supplementary Information for further details). Figures 4(a) and 4(b) show, respectively, the simulated bottom conduction-band energies $E_C$ and the extracted out-of-plane electric field distributions in the InAs nanosheet with a carrier density $n$ ~ 1.34×10$^{12}$ cm$^{-2}$ (corresponding to the nanosheet conductance $G$ ~ 12$e^2$/h) for a few experimentally applied dual-gate voltages $V_D$. It is seen that at $V_D$ = –3.36 V, the simulated $E_C$ is strongly downward tilted, leading to the presence of strong, positive, out-of-plane electric fields and thus the presence of strong Rashba SOI in the InAs nanosheet. As $V_D$ decreases, the simulated $E_C$ becomes less downward tilted and thus the out-of-plane field strength is reduced, leading to a decrease in the Rashba SOI strength. At the point when $V_D$ = –6.28 V, the simulation gives a flat $E_C$ profile and thus zero electric field inside the InAs nanosheet. As a result, no Rashba SOI should be present in the InAs nanosheet. When continuously decreasing $V_D$ to $V_D = -7.62$ V, the simulated $E_C$ becomes upward tilted, leading to the presence of negative, out-of-plane electric fields in the InAs nanosheet and thus a reappearance of Rashba SOI. Nevertheless, at $V_D = -7.62$ V, the field strength in the nanosheet is in general less than it at $V_D = -3.36$ V. Thus, the Rashba SOI at $V_D = -7.62$ V is weaker when comparing to the case of $V_D = -3.36$ V. The evolution of the Rashba SOI in the InAs nanosheet with decreasing $V_D$, inferred above from the simulation, is consistent with our observation shown in Fig. 3. Thus, tuning spin-orbit length $L_{SO}$ in the InAs nanosheet from a short to a long and then back to a short value originates from tuning Rashba SOI by the dual gate. Note that in the experiment, at $V_D = -6.28$ V, $L_{SO}$ does not go to infinity. This is because a weak, but finite, dual-gate voltage independent SOI term is present in the InAs nanosheet. This term is dominantly of the Dresselhaus type as a result of breaking bulk inversion symmetry in the InAs compound crystal.

In many cases, it could be convenient to characterize the Rashba SOI strength using the Rashba spin-orbit coupling constant $\alpha$ defined as $\alpha = r_R E$, where $E$ is an average strength of the out-of-plane electric field $\boldsymbol{E}$ in the InAs nanosheet and $r_R$ is a material-specific, Fermi-level dependent prefactor found in the Rashba SOI Hamiltonian $H_R = r_R \boldsymbol{\sigma} \cdot \boldsymbol{k} \times \boldsymbol{E}$.[23] Since the Rashba spin-orbit precession length is given by $L_{SO}^R = \frac{\hbar^2}{m^* \alpha} = \frac{\hbar^2}{m^* r_R E}$ (where $m^*$ is the electron effective mass), one can write, to a good approximation, the measured spin-orbit length $L_{SO}$ as $\frac{1}{L_{SO}^2} = \frac{1}{(L_{SO}^R)^2} + C_0 = (\frac{m^* r_R}{\hbar^2})^2 E^2 + C_0$, where $C_0$ takes into account all the electric field-independent contributions to the spin processions in the InAs nanosheet.[12] To estimate the Rashba coupling constant $\alpha$ in the InAs nanosheet, we evaluate the average out-of-plane electric field strengths $E$ at different $V_D$ from the simulated results shown in Fig. 4(b) and plot the extracted $\frac{1}{L_{SO}^2}$ as a function of $E^2$ in Fig. 4(c) (data points). The dashed line in Fig. 4(c) is a linear fit to the data points, whose slope gives a value of $r_R = 51.93$ e·nm². Figure 4(d) shows the extracted values of $|\alpha|$ at the experimentally applied values of $V_D$. It is seen that $|\alpha|$ can be tuned from zero to a value of ~0.36 eVÅ for the InAs nanosheet in Device 1 at the conductance of $G \sim 12e^2/h$ (with the corresponding carrier density of $n \sim 1.34 \times 10^{12}$ cm⁻² in the nanosheet). Note that $|\alpha|$ is plotted in Fig. 4(d) and in fact the Rashba coupling constant $\alpha$ in the InAs nanosheet of Device 1 can be tuned from a large negative value to a large positive value. The value of $C_0 \sim 11$ μm⁻² could also be extracted from the fit shown in Fig. 4(c). If we assume that the remaining field-independent contribution to the spin procession is dominantly from the Dresselhaus SOI in the InAs nanosheet, the Dresselhaus spin-orbit precession length can then be extracted as $L_{SO}^{D} \sim 0.30\ \mu m$ at the nanosheet conductance of $G \sim 12e^2/h$. Similar analyses have also been carried out for Device 1 at the conductance values of $G \sim 7e^2/h$ and $G \sim 4e^2/h$, and the same large tunability in the Rashba coupling constant $\alpha$ has been obtained (see Section V and Fig. S5 in Supplementary Information). In comparison, we note that the tunable Rashba coupling constants of $\alpha \sim 0.01$–$0.1$ eVÅ were observed in the InGaAs quantum well and InAs quantum well structures.[32-35]

Finally, we discuss the influence of temperature on the quantum transport characteristics of the InAs nanosheet device (Device 1). Figure 5(a) shows the low-field magnetoconductance $\Delta G$ measured for Device 1 at $V_{BG} = -6$ V and $V_{TG} = -0.6$ V at temperatures $T = 2$ to 20 K. At $T = 2$ K, the measured $\Delta G$ displays clearly a peak-like structure near zero magnetic field, indicating that the electron transport in the nanosheet exhibits dominantly the WAL characteristic. As $T$ increases from 2 to 20 K, the WAL characteristic is gradually suppressed. Again, the measured $\Delta G$ data at different temperatures are fitted to the HLN formula [Eq. (1)]. The solid lines in Fig. 5(a) are the fitting results. Figure 5(b) displays the extracted values of $L_\varphi$, $L_{SO}$, and $L_e$ as a function of $T$ from the fits. It is generally seen that both $L_{SO}$ and $L_e$ show a weak temperature dependence. However, $L_\varphi$ is strongly dependent on temperature—it decreases from ~235 to ~112 nm with increasing $T$ from 2 to 20 K. The extracted $L_\varphi$ data at different temperatures can be fitted by a power law of $L_\varphi \sim T^{-0.37}$ [see the solid line in Fig. 5(b)]. This $L_\varphi \sim T^{-0.37}$ dependence suggests that the dephasing of electrons in the InAs nanosheet is dominantly caused by electron-electron interactions with small energy transfer (the Nyquist scattering mechanism).[36] Note that the width of the InAs nanosheet conduction channel in the device is only slightly larger than the extracted phase coherence length ($W \sim 410$ nm vs. $L_\varphi \sim 120$–$230$ nm), the observed $L_\varphi \sim T^{-0.37}$ falls in a result lying between the one-dimensional ($T^{-1/3}$) and the 2D ($T^{-1/2}$) regime.[37]

In summary, a dual-gate field-effect device has been made from an epitaxially grown free-standing InAs nanosheet and studied by transport measurements at low temperatures. It is shown that both the carrier density in the InAs nanosheet and the potential difference across the nanosheet can be tuned independently with use of the dual gate. The magnetoconductance measurements show that the electron transport in the nanosheet can be tuned from the WAL to the WL and then back to the WAL characteristic by sweeping the dual-gate voltage along a constant nanosheet channel conductance contour. Thus, an efficient tuning of the Rahsba SOI in the InAs nanosheet can be achieved at a constant carrier density. The extracted spin-obit length at a constant carrier density is found to exhibit a peak value at which the Rashba SOI is suppressed and the spin relaxation by the Dresselhaus SOI can be revealed. Magnetotransport measurements have also been carried out for a dual-gate device made from another InAs nanosheet and similar results are found (see Section VI in Supplementary Information). We have also performed the energy band diagram simulations for the layer structure of a dual-gate InAs nanosheet device at experimental conditions and the obtained results support our experimental observations.

## AUTHOR CONTRIBUTIONS

H. Q. Xu conceived and supervised the project. F. Fan fabricated the devices, performed the transport measurements and carried out the simulations for energy band diagrams with support from Y. Chen. D Pan and J. Zhao grew the materials and analyzed the material crystal structures. F. Fan and H. Q. Xu analyzed the transport measurement data and the simulation results. F. Fan and H. Q. Xu wrote the manuscript with contributions from all the authors.

## CONFLICTS OF INTEREST

There are no conflicts of interests to declare.

## ACKNOWLEDGMENTS

This work was supported by the Ministry of Science and Technology of China through the National Key Research and Development Program of China (Grant Nos. 2017YFA0303304 and 2016YFA0300601), the National Natural Science Foundation of China (Grant Nos. 92165208, 92065106, 61974138, 11874071, 91221202, and 91421303), and the Beijing Academy of Quantum Information Sciences (Grant No. Y18G22). D. P. also acknowledges support from the Youth Innovation Promotion Association, Chinese Academy of Sciences (2017156).

**FIGURE CAPTIONS**

**FIG. 1.** (a) SEM image (side view) of MBE-grown free-standing InAs nanosheets on a Si (111) substrate. (b) HRTEM image of an InAs nanosheet taken from the substrate shown in (a). The inset is the corresponding SAED pattern recorded along the $[2\bar{1}\bar{1}0]$ axis. (c) SEM image (top view, in false color) of a fabricated dual-gate field-effect InAs nanosheet device (Device 1) studied in this work and measurement circuit setup. (d) Schematic for the layer structure of the device in (c).

**FIG. 2.** Conductance $G$ measured for Device 1 as a function of bottom-gate voltage $V_{BG}$ and top-gate voltage $V_{TG}$ at $T = 2$ K. The pink, purple and orange solid lines represent the constant conductance $G$ contours of ~12, ~7 and ~4e$^2$/h, respectively.

**FIG. 3.** (a) Low-field magnetoconductance $\Delta G$ measured for Device 1 at $T = 2$ K and at dual-gate voltages $V_D$ taken along the contour of $G \sim 12$e$^2$/h. Here, $\Delta G = G(B) - G(B = 0)$, $V_D = V_{BG} - V_{TG}$, and the applied magnetic field $B$ is perpendicular to the InAs nanosheet plane. The bottom black data points display the measured $\Delta G$ at $V_D = -3.36$ V and the $\Delta G$ data measured at other values of $V_D$ are successively vertically offset by 0.02e$^2$/h for clarity. The gray solid lines are the results of the fits of the measured data to the HLN theory [Eq. (1)]. (b) Characteristic transport lengths $L_\varphi$, $L_{SO}$, and $L_e$ extracted from the fits in (a) as a function of $V_D$.

**FIG. 4.** (a) Simulated conduction band minimums $E_C$ in the InAs nanosheet with a fixed value of $n \sim 1.34 \times 10^{12}$ cm$^{-2}$ (corresponding to $G \sim 12$e$^2$/h) for Device 1 at different $V_D$. (b) Corresponding effective out-of-plane electric fields in the InAs nanosheet. (c) Extracted $\frac{1}{L_{SO}^2}$ versus evaluated $E^2$ in the InAs nanosheet. Here, $E$ represents the average out-of-plane electric field strength in the nanosheet and the dashed line is a linear fit to the data points. (d) Estimated values of the Rashba spin-orbit coupling constant $|\alpha|$ in the InAs nanosheet of Device 1 at a few experimentally applied values of $V_D$.

**FIG. 5.** (a) Low-field magnetoconductance $\Delta G$ measured for Device 1 at $V_{BG} = -6$ V and $V_{TG} = -0.6$ V and at different temperatures $T$. The bottom red data points display the measured $\Delta G$ data at $T = 2$ K and the $\Delta G$ data measured at other temperatures are successively vertically offset by 0.01e$^2$/h for clarity. The gray solid lines are the results of the fits of the measured $\Delta G$ data to the HLN theory [Eq. (1)]. (b) Characteristic transport lengths $L_\varphi$, $L_{SO}$, and $L_e$ extracted from the fits in (a) as a function of $T$. The black solid line is a power-law fit of the extracted values of $L_\varphi$, showing $L_\varphi \sim T^{-0.37}$.

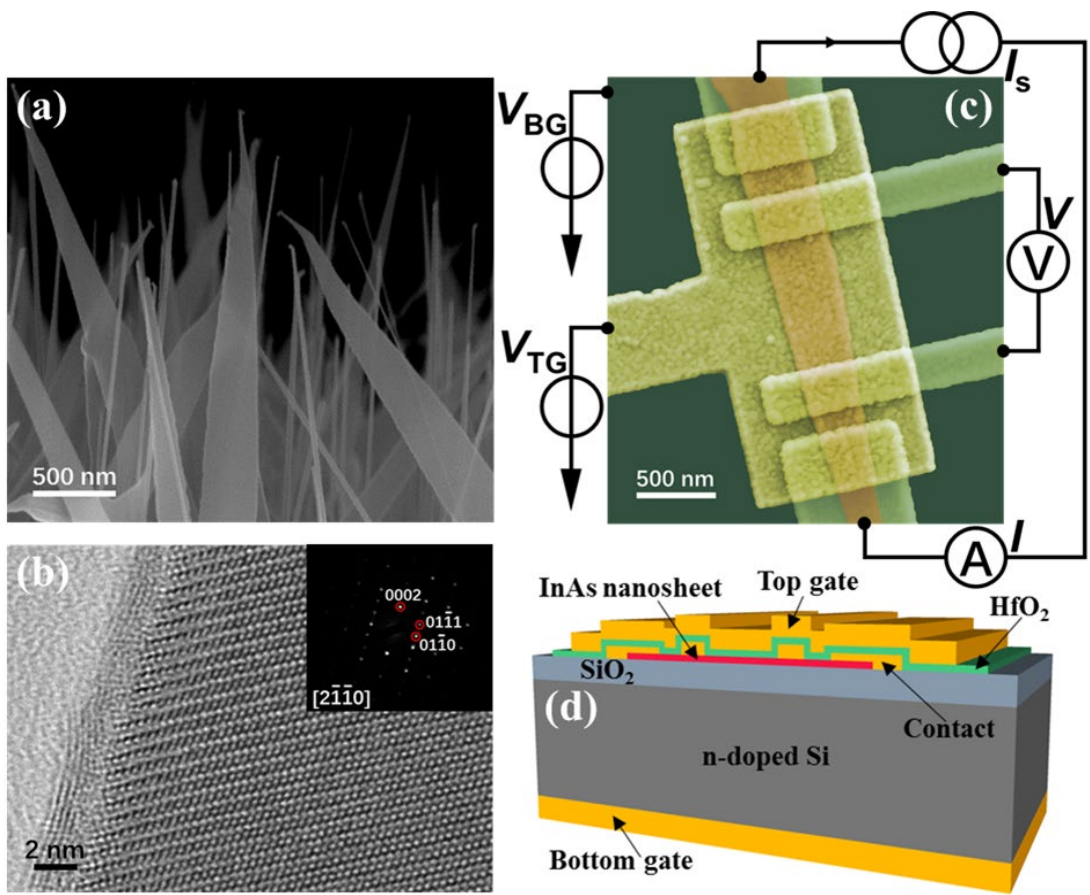

**Figure 1, Furong Fan *et al*.**

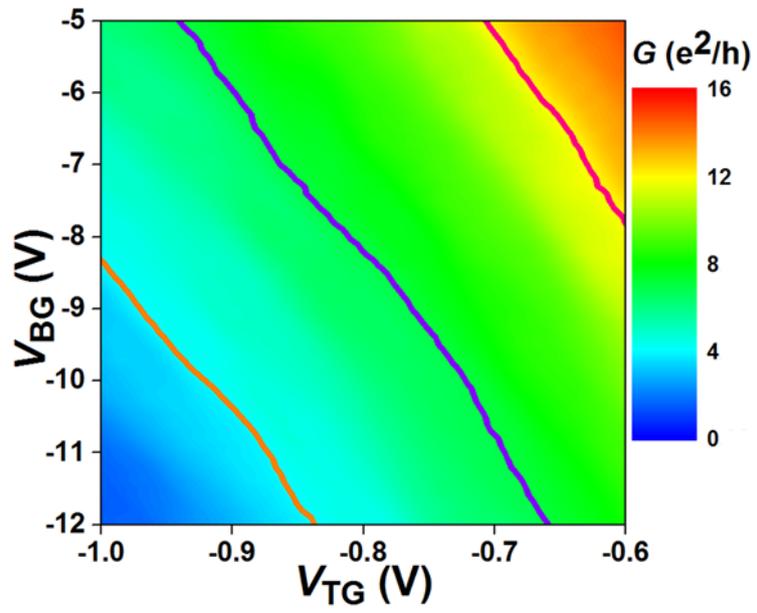

**Figure 2, Furong Fan** *et al.*

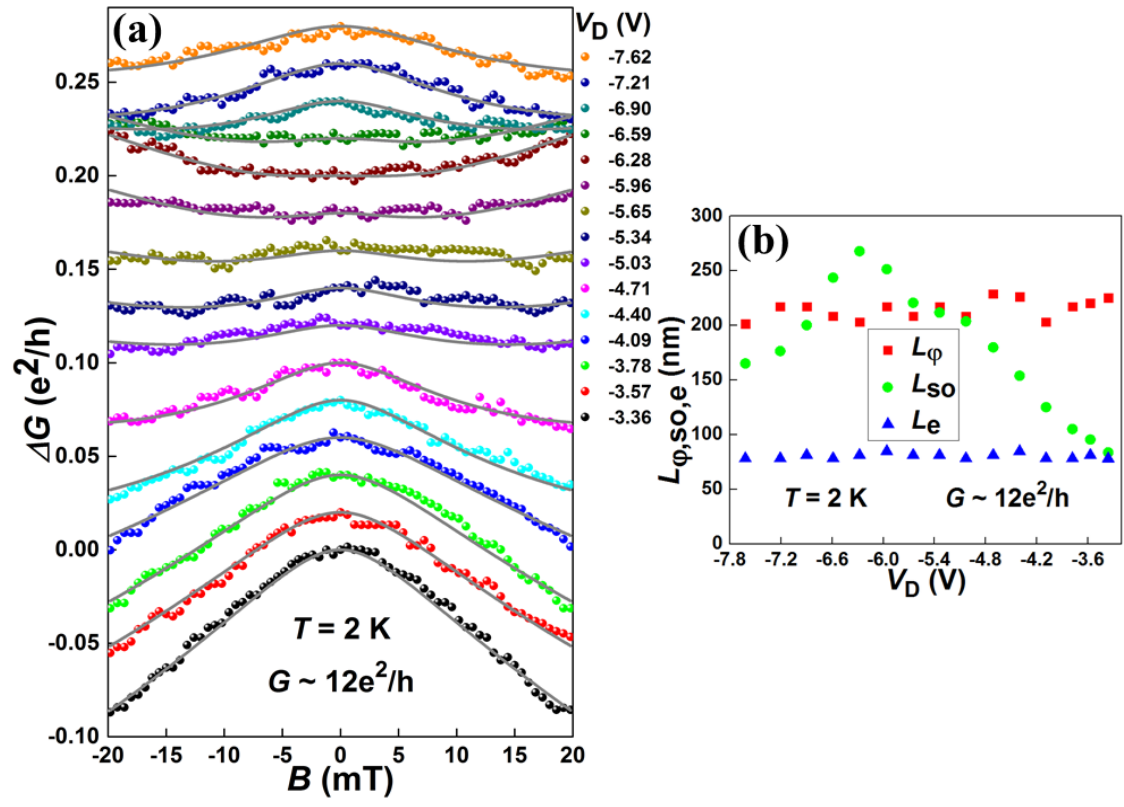

Figure 3, Furong Fan *et al.*

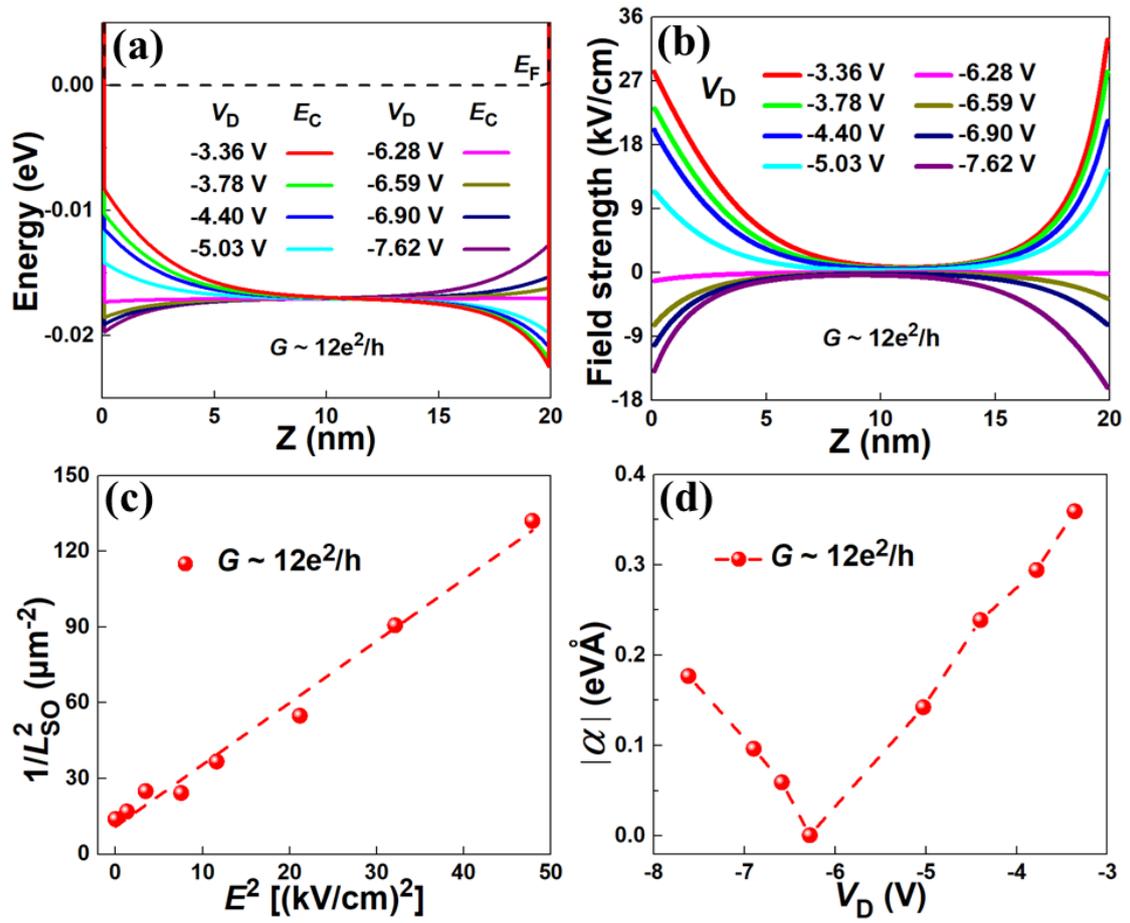

Figure 4, Furong Fan *et al*.

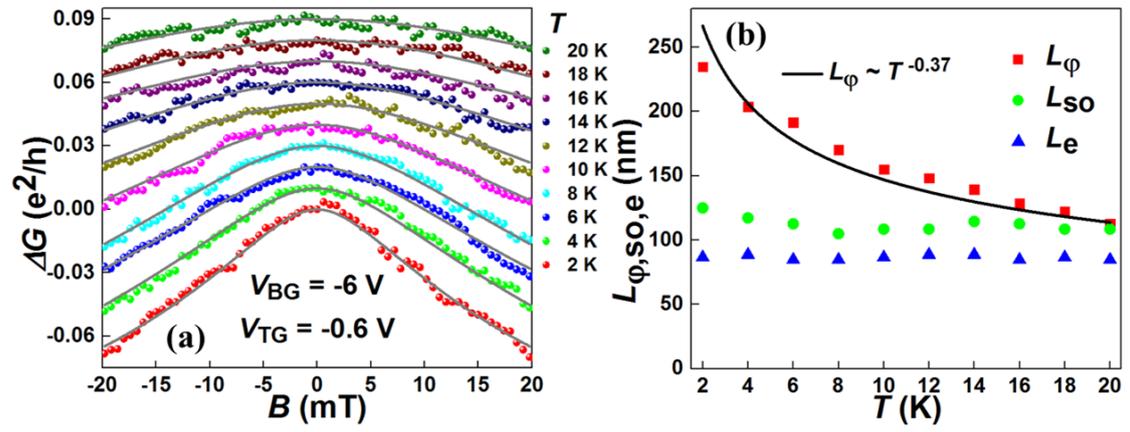

Figure 5, Furong Fan *et al*.

**Supplementary Information for**

# Electrically tunable spin-orbit interaction in an InAs nanosheet

Furong Fan[1], Yuanjie Chen[1], Dong Pan[2], Jianhua Zhao[2], and H. Q. Xu[1,3,*]

[1]*Beijing Key Laboratory of Quantum Devices, Key Laboratory for the Physics and Chemistry of Nanodevices and School of Electronics, Peking University, Beijing 100871, China*

[2]*State Key Laboratory of Superlattices and Microstructures, Institute of Semiconductors, Chinese Academy of Sciences, P.O. Box 912, Beijing 100083, China*

[3]*Beijing Academy of Quantum Information Sciences, Beijing 100193, China*

*Corresponding author; email: hqxu@pku.edu.cn

(Date: April 30, 2022)

# Contents

In this Supplementary Information, we provide, in Sections I, II, III, IV, and V, additional data for the dual-gate field-effect InAs nanosheet device (i.e., Device 1) studied in the main article. In Section VI, we provide the results of measurements obtained from a similar dual-gate field-effect device (named as Device 2) made from another epitaxially grown InAs nanosheet.

**Section I.** Carrier density ($n$), Fermi wavelength ($\lambda_F$), and mean free path ($L_e$) in the InAs nanosheet extracted from the device gate-transfer characteristic measurements

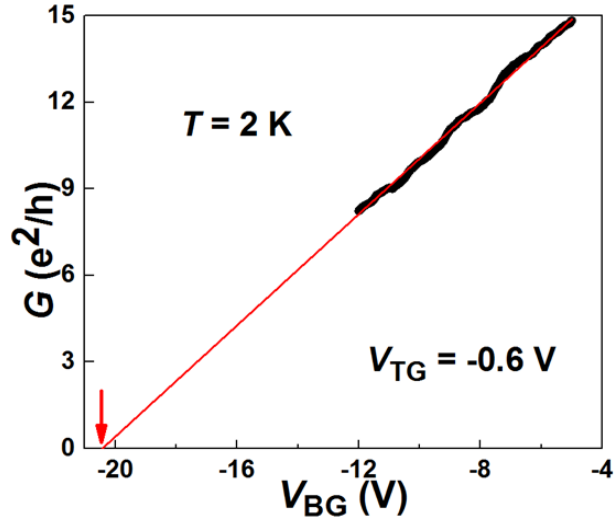

**FIG. S1.** Conductance $G$ (black line) measured for Device 1 as a function of $V_{BG}$ at fixed $V_{TG} = -0.6$ V and $T = 2$ K. The red line is a linear fit to the measured data. The red arrow marks the intersection of the fitting line and the $V_{BG}$ axis, i.e., the bottom-gate threshold voltage of $V_{BG}^{th} = -20.2$ V at $V_{TG} = -0.6$ V.

In this Section, we describe how the carrier density ($n$), Fermi wavelength ($\lambda_F$), and mean free path ($L_e$) in the InAs nanosheet at different conductance values are extracted from the gate-transfer characteristic measurements. The carrier density $n$ in the InAs nanosheet can be estimated from the measured bottom-gate transfer characteristics at a fixed top-gate voltage based on a parallel capacitor model of $n = \frac{C_{BG}(V_{BG} - V_{BG}^{th})}{eA}$ with $e$ being the elementary charge, $A$ the area of the InAs nanosheet conduction channel, $C_{BG}$ the bottom-gate capacitance to the nanosheet channel, $V_{BG}$ the applied bottom-gate voltage, and $V_{BG}^{th}$ the bottom-gate threshold voltage at which the nanosheet channel starts to open for conduction. The capacitance $C_{BG}$ can be calculated from $C_{BG} = \frac{\varepsilon_0 \varepsilon_r A}{d}$, where $\varepsilon_0$ is the permittivity of vacuum, $\varepsilon_r$ is the relative permittivity of SiO$_2$, and $d$ is the thickness of SiO$_2$. In this work for Device 1, using $\varepsilon_0 \sim 8.85 \times 10^{-12}$ F/m, $\varepsilon_r \sim 3.9$, $d \sim 200$ nm and $A \sim 0.34$ µm$^2$, $C_{BG}$ can be estimated out as $5.87 \times 10^{-17}$ F. The threshold voltage $V_{BG}^{th}$ can be extracted from the fit of the measured $G \sim V_{BG}$ characteristics at a fixed $V_{TG}$ to $G = \frac{\mu}{L^2} C_{BG} \left( V_{BG} - V_{BG}^{th} \right)$, where $\mu$ is the field-effect mobility of electrons in the nanosheet. Figure S1 shows the measured $G \sim V_{BG}$ characteristics of Device 1 (black curve) at $V_{TG} = -0.6$ V and $T = 2$ K and the fitting result (red curve). From the fit, a threshold voltage of $V_{BG}^{th} \sim -20.2$ V, as marked by a red arrow in Fig. S1, and a field-effect mobility of $\mu \sim 4460$ cm$^2$/Vs are extracted. Now, the carrier density $n$ in the InAs nanosheet at different conductance values can be estimated. For example, at the channel conductance of $G \sim 12e^2/h$, the carrier density can be estimated out as $n \sim 1.34 \times 10^{12}$ cm$^{-2}$ by taking $V_{TG} = -0.6$ V and $V_{BG} = -7.8$ V (cf. the dual-gate transfer characteristic measurements shown in Fig. 2 of the main article and the bottom-gate transfer characteristic measurements shown in Fig. S1). The Fermi wavelength can be estimated from $\lambda_F = \sqrt{2\pi/n}$, giving $\lambda_F \sim 22$ nm at $n \sim 1.34 \times 10^{12}$ cm$^{-2}$. The corresponding mean free path in the nanosheet is $L_e \sim 85$ nm, obtained from $L_e = \frac{\hbar \mu}{e} \sqrt{2\pi n}$ (where $\hbar = \frac{h}{2\pi}$ is the reduced Planck constant). Similarly, the values of $n$, $\lambda_F$, and $L_e$ can be extracted from the measured gate-transfer characteristics for the InAs nanosheet at other nanosheet channel conductance values. Table S1 lists the extracted values of $n$, $\lambda_F$, and $L_e$ for the InAs nanosheet in Device 1 at the three channel conductance $G$ values of $\sim 12$, $\sim 7$ and $\sim 4e^2/h$.

**TABLE S1.** Extracted carrier densities $n$, Fermi wavelengths $\lambda_F$, and mean free paths $L_e$ for the InAs nanosheet in Device 1 at three channel conductance $G$ values of $\sim 12$, $\sim 7$ and $\sim 4e^2/h$.

| Conductance $G$ (e$^2$/h) | Carrier Density $n$ (cm$^{-2}$) | Fermi Wavelength $\lambda_F$ (nm) | Mean Free Path $L_e$ (nm) |
|---|---|---|---|
| 12 | $1.34 \times 10^{12}$ | 22 | 85 |
| 7 | $1.07 \times 10^{12}$ | 24 | 56 |
| 4 | $5.8 \times 10^{11}$ | 33 | 41 |

**Section II.** Tuning of the low-field magnetotransport characteristics of the InAs nanosheet in Device 1 by the bottom and the top gate individually

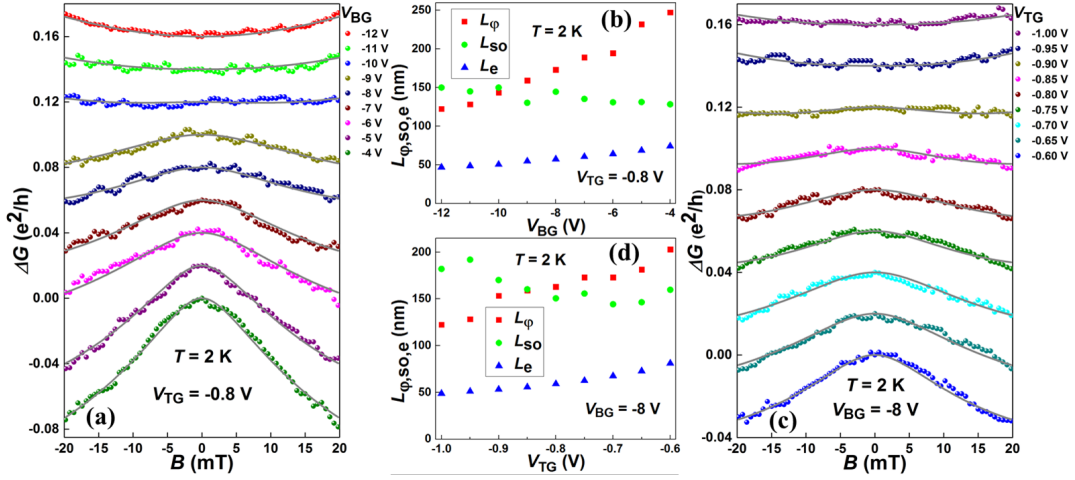

**FIG. S2.** (a) Low-field magnetoconductance $\Delta G$ measured for Device 1 at different $V_{BG}$, with $V_{TG}$ fixed at $V_{TG} = -0.8$ V, and at $T = 2$ K. The bottom green data points display the measured $\Delta G$ data at $V_{BG} = -4$ V and the data points measured at all other values of $V_{BG}$ are successively vertically offset by $0.02e^2/h$ for clarity. The gray solid lines are the results of the fits of the measured $\Delta G$ data to the HLN theory [Eq. (1) in the main article]. (b) Characteristic transport lengths $L_{\varphi}$, $L_{SO}$ and $L_e$ extracted as a function of $V_{BG}$ from the fits shown in (a). (c) Low-field magnetoconductance $\Delta G$ measured for Device 1 at different $V_{TG}$, with $V_{BG}$ fixed at $V_{BG} = -8$ V, and at $T = 2$ K. The bottom blue data points display the measured $\Delta G$ data at $V_{TG} = -0.60$ V and the data points measured at all other values of $V_{TG}$ are successively vertically offset by $0.02e^2/h$ for clarity. The gray solid lines are again the results of the fits of the measured $\Delta G$ data to the HLN theory [Eq. (1) in the main article]. (d) Characteristic transport lengths $L_{\varphi}$, $L_{SO}$, and $L_e$ extracted as a function of $V_{TG}$ from the fits shown in (c).

In this Section, we provide additional measurement data for the influences of the bottom- and top-gate voltages on the low-field magnetotransport characteristics of the InAs nanosheet in Device 1. Figure S2(a) shows the measured low-field magnetoconductance $\Delta G$ of Device 1 as a function of bottom-gate voltage $V_{BG}$ at a fixed top-gate voltage of $V_{TG} = -0.8$ V and at $T = 2$ K. Figure S2(c) shows the measured $\Delta G$ of Device 1 as a function of $V_{TG}$ at fixed $V_{BG} = -8$ V and at $T = 2$ K. It is seen that both $V_{BG}$ and $V_{TG}$ can modulate the low-field magnetotransport characteristics of the InAs nanosheet effectively, turning the electron transport in the InAs nanosheet from the weak localization (WL) regime to the weak antilocalization (WAL) regime. The solid lines in Figs. 2(a) and 2(c) represent the fits of the measurements to the theory of Hikami, Larkin, and Nagaoka (HLN) [i.e., Eq. (1) in the main article]. Figures S2(b) and S2(d) display the extracted values of $L_\varphi$, $L_{SO}$, and $L_e$ from the fits shown in Figs. S2(a) and S2(c), respectively. Here, it is seen that with varying $V_{BG}$ or $V_{TG}$ alone, $L_{SO}$ does not get largely changed, though an effective change in carrier density $n$ (not shown here) and in phase coherence length $L_\varphi$ can be achieved. In fact, it is the efficient change in phase coherence length $L_\varphi$ with change in $V_{BG}$ or in $V_{TG}$ that leads to a crossover between $L_\varphi$ and $L_{SO}$ and thus a transition from the WL to the WAL regime in the electron transport in the InAs nanosheet. The above results provide a clear evidence that it is difficult to efficiently tune the spin-orbit interaction (SOI) in the InAs nanosheet by a single gate. Thus, in the main article, a dual-gate device setup is developed in order to achieve an efficient tuning of the electric field and thus the Rashba SOI in the InAs nanosheet without a change in carrier density.

**Section III.** Simulations for the energy band diagrams of the HfO$_2$/InAs/SiO$_2$ layer structure in Device 1

To get the physical insights into the experimental results presented in the main article, we carry out the simulations for the energy band diagrams of the HfO$_2$/InAs/SiO$_2$ layer structure in the studied dual-gate InAs nanosheet device (Device 1). In the simulations, Poisson's equation is solved in compliance with the boundary conditions set by the experiments. The material parameters of HfO$_2$, InAs, and SiO$_2$ used in the energy band diagram simulations are given in Table S2. Here, in the simulations, the three layers of HfO$_2$, InAs, and SiO$_2$ are assumed to be infinite in the lateral dimensions and thus the electrostatics of the HfO$_2$/InAs/SiO$_2$ layered system can be described by an effective one-dimensional Poisson's equation of $\varepsilon \nabla_z^2 \varphi(z) = -\rho(z)$, where $\varepsilon$ is the dielectric constant of the material, $\varphi(z)$ is the electric potential, and $\rho(z)$ is the charge density. However, the layer thicknesses of HfO$_2$, InAs, and SiO$_2$ are taken as 20 nm, 20 nm, and 200 nm, respectively, which are in close correspondence to the structure of Device 1. The charge density $\rho(z)$ is given by $\rho(z) = e[p(z) - n(z) + N_d(z) - N_a(z)]$, where $p(z)$, $n(z)$, $N_d(z)$, and $N_a(z)$ are the hole density, electron density, donor concentration and acceptor concentration, respectively. The energies of the conduction band minimum $E_C(z)$ and the valence band maximum $E_V(z)$ are given by $E_C(z) = -[e\varphi(z) + \chi]$ and $E_V(z) = -[e\varphi(z) + \chi + E_g]$, where $\chi$ and $E_g$ are the electron affinity and bandgap of the material. We require that the electron Fermi level $E_F$ is continuous at the interface of two different materials and that the simulated charge density in the InAs material is equal to the charge density in the InAs nanosheet extracted from the gate transfer characteristic measurements.

**TABLE S2.** Material parameters of HfO$_2$, InAs, and SiO$_2$ employed in the simulations for the energy band diagrams of the HfO$_2$/InAs/SiO$_2$ layer structure in Device 1.

| Material | Bandgap (eV) | Dielectric Constant ($\varepsilon_0$) | Electron Effective Mass ($m_e$) | Electron Affinity (eV) |
|----------|--------------|---------------------------------------|----------------------------------|------------------------|
| HfO$_2$  | 5.8          | 9.1                                   | 0.42                             | 2.8                    |
| InAs     | 0.354        | 15.15                                 | 0.023                            | 4.9                    |
| SiO$_2$  | 8.95         | 3.9                                   | 0.427                            | 0.75                   |

Figure S3(a) shows the simulated energy band diagram of the HfO$_2$/InAs/SiO$_2$ layer structure in Device 1 at $V_{BG} = V_{TG} = 0$ V ($V_D = 0$ V) and Fig. S3(b) is a zoom-in plot of the energy band diagram in the InAs nanosheet. It should be noted that even in the case of $V_{BG} = V_{TG} = 0$ V, the band structure in the InAs nanosheet is asymmetric, leading to a structural asymmetric and thus the presence of a Rashba SOI in the InAs nanosheet.

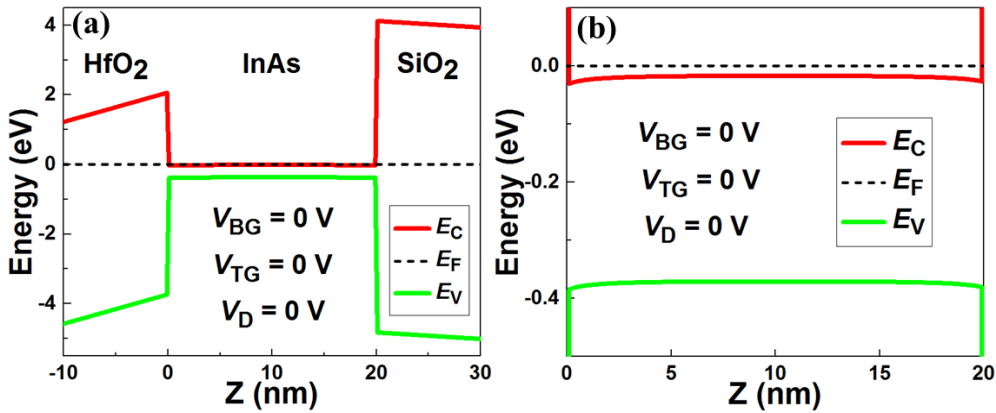

**FIG. S3.** (a) Simulated energy band diagram of the HfO$_2$/InAs/SiO$_2$ layer structure in Device 1 at $V_{BG} = V_{TG} = 0$ V ($V_D = 0$ V). (b) Zoom-in plot of the energy band diagram in the InAs nanosheet.

**Section IV.** Low-field magnetotransport characteristics measured for Device 1 as a function of dual-gate voltage $V_D$ taken along the constant conductance contours of $G \sim 7e^2/h$ and $G \sim 4e^2/h$

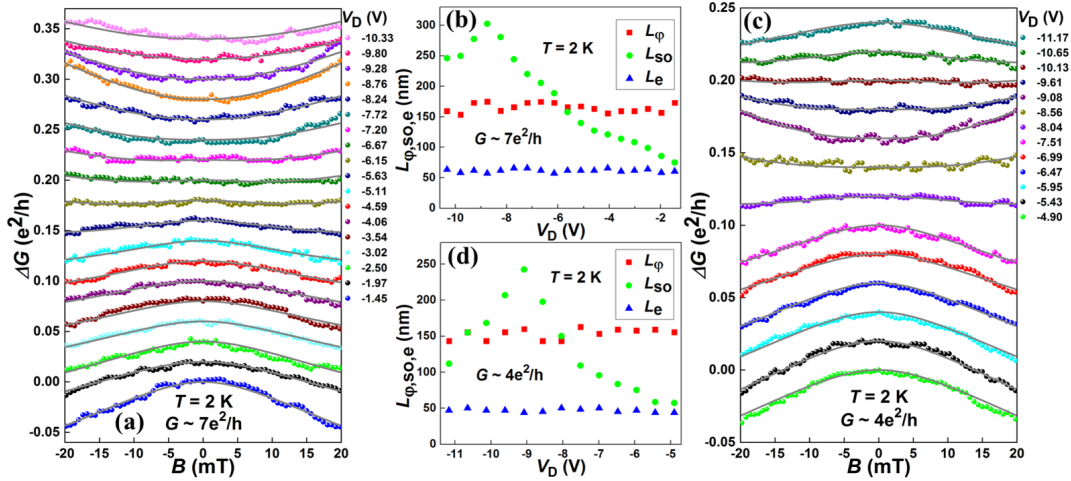

**FIG. S4.** (a) Low-field magnetoconductance $\Delta G$ measured for Device 1 at $T = 2$ K and at a series of dual-gate voltages $V_D$ taken along the constant conductance contour of $G \sim 7e^2/h$. The bottom blue data points display the measured $\Delta G$ data at $V_D = -1.45$ V and the $\Delta G$ data measured at all other $V_D$ values are successively vertically offset by $0.02e^2/h$ for clarity. The gray solid lines are the results of the fits of the measurements to the HLN theory [Eq. (1) in the main article]. (b) Characteristic transport lengths $L_\varphi$, $L_{SO}$, and $L_e$ as a function of $V_D$, extracted from the fits in (a). (c) The same as in (a) but for the measurements at different $V_D$ values taken along the constant conductance contour of $G \sim 4e^2/h$. The bottom green data points display the measured $\Delta G$ data at $V_D = -4.90$ V and the $\Delta G$ data measured at all other $V_D$ values are successively vertically offset by $0.02e^2/h$ for clarity. Again, the gray solid lines are the results of the fits of the measurements to the HLN theory [Eq. (1) in the main article]. (d) Characteristic transport lengths $L_\varphi$, $L_{SO}$, and $L_e$ as a function of $V_D$, extracted from the fits in (c).

In this section, we provide additional data measured for the low-field magnetotransport characteristics of Device 1 at the constant conductance values of $\sim 7$ and $\sim 4e^2/h$. Figures S4(a) and S4(c) show the low-field magnetoconductance $\Delta G$ measured for Device 1 at $T=2$ K as a function of dual-gate voltage $V_D$ taken along the constant conductance contours of $G \sim 7e^2/h$ and $G \sim 4e^2/h$, respectively. The solid lines in the figures are the results of the fits of the measurement data to the HLN theory [Eq.

(1) in the main article]. Figures S4(b) and S4(d) show the extracted characteristic transport lengths $L_\varphi$, $L_{SO}$ and $L_e$ from the fits shown in Figs. S4(a) and S4(c), respectively. Here, both $L_\varphi$ and $L_e$ are seen to show a weak dependence on $V_D$. However, with decreasing $V_D$, the extracted $L_{SO}$ undergoes a dramatic change—it first increases quickly and then falls down. This efficient modulation of $L_{SO}$ in the InAs nanosheet at different constant carrier densities (revealed by staying at different constant conductance values) by dual-gate voltage $V_D$ reveals again that an SOI of the Rashba type is present in the InAs nanosheet and is greatly tunable with use of the dual gate, in full consistence with the results presented in the main article.

**Section V.** Rashba spin-orbit coupling constant $\alpha$ and its extraction for the InAs nanosheet in Device 1 at the conductance values of $G \sim 7e^2/h$ and $\sim 4e^2/h$.

We have extracted the spin-orbit lengths $L_{SO}$ in the InAs nanosheet of Device 1 at three fixed conductance values of $G \sim 12e^2/h$, $\sim 7e^2/h$ and $\sim 4e^2/h$ from the low-field magnetotransport measurements shown in Figs. 3 and S4. The efficient tuning of $L_{SO}$ at a constant conductance with use of the dual gate demonstrates that the tuned SOI in the InAs nanosheet is dominantly of Rashba type. However, all kinds of SOIs including e.g., Rashba type, Dresselhaus type, and other high-order types, in the nanosheet would contribute to spin relaxation rate $\frac{1}{\tau_{SO}} \propto \frac{1}{L_{SO}^2}$ and thus affect $L_{SO}$. In this section, we describe how the Rashba SOI strengths in the InAs nanosheet of Device 1 are estimated and provide additional data for the results of simulations obtained at the constant conductance values of $G \sim 7e^2/h$ and $\sim 4e^2/h$.

The origin of the Rashba SOI is the presence of an effective out-of-plane electric field induced by structural inversion asymmetry in the InAs nanosheet. To the lowest-order approximation, the Rashba SOI can be described by the Hamiltonian of $H_R = r_R \boldsymbol{\sigma} \cdot \boldsymbol{k} \times \boldsymbol{E}$, where $\boldsymbol{\sigma}$, $\boldsymbol{k}$, and $\boldsymbol{E}$ are the Pauli spin matrices, wave vector, and out-of-plane electric field, respectively. The Rashba SOI strength can be conveniently characterized by the Rashba coupling constant $\alpha = r_R E$, where $r_R$ is a material-specific, Fermi-level dependent prefactor and $E$ is the average strength of the out-of-plane electric field $\boldsymbol{E}$ in the InAs nanosheet. The Rashba spin-orbit precession length is related to the Rashba coupling constant $\alpha$ via $L_{SO}^R = \hbar^2/m^*\alpha$, with $m^*$ being the electron effective mass. Because the total spin relaxation time rate can be written as $\frac{1}{\tau_{SO}} = \frac{1}{\tau_R} +$ $\cdots$, where $\frac{1}{\tau_R}$ is the term originated from the Rashba SOI, we can get $\frac{1}{L_{SO}^2} = \left(\frac{m^* r_R}{\hbar^2}\right)^2 E^2 + C_0$, where $C_0$ is a field-independent constant.

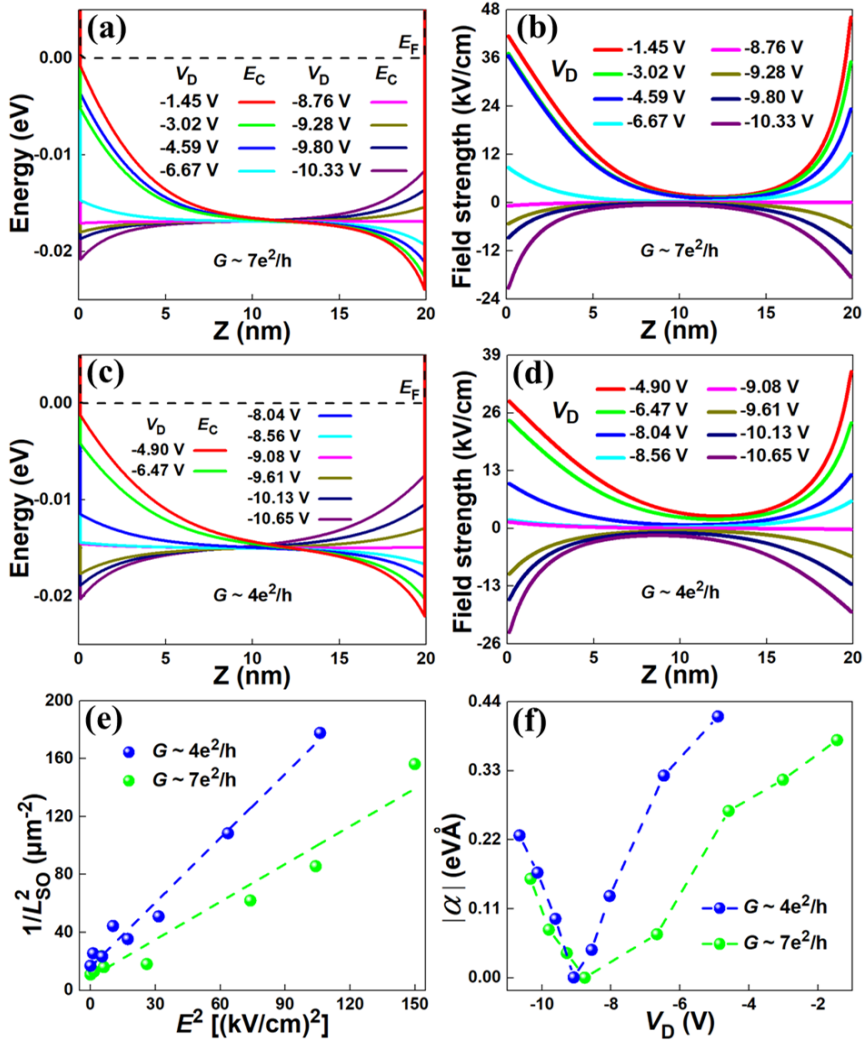

**FIG. S5.** (a) Simulated conduction band minimums $E_C$ in the InAs nanosheet with a fixed value of $G \sim 7e^2/h$ (corresponding to $n \sim 1.07 \times 10^{12}$ cm$^{-2}$ in the InAs nanosheet) for Device 1 at different dual-gate voltages $V_D$. (b) Electric field distributions in the InAs nanosheet derived from the simulated conduction band structures shown in (a). (c) and (d) The same as in (a) and (b), but for the device at $G \sim 4e^2/h$ (corresponding to $n \sim 5.8 \times 10^{11}$ cm$^{-2}$ in the nanosheet). (e) Extracted $\frac{1}{L_{SO}^2}$ versus calculated $E^2$ in the InAs nanosheet. Here, $E$ represents the average electric field strength in the nanosheet. Green and blue dots are the data points obtained at $G \sim 7e^2/h$ and $G \sim 4e^2/h$, respectively. The dashed lines are the linear fits to the data. (f) Estimated values of the Rashba spin-orbit coupling constant $|\alpha|$ in the InAs nanosheet of Device 1 at a few experimentally applied values of $V_D$.

Figures S5(a) and S5(c) show the simulated energy band diagrams in the InAs nanosheet of Device 1 at different applied $V_D$ for the two fixed conductance values of $G \sim 7e^2/h$ and $\sim 4e^2/h$ (corresponding to $n \sim 1.07 \times 10^{12}$ cm$^{-2}$ and $n \sim 5.8 \times 10^{11}$ cm$^{-2}$ in the InAs nanosheet), respectively. Figures S5(b) and S5(d) show the corresponding effective out-of-plane electric field distributions in the InAs nanosheet, from which the average electric field strength $E$ in the InAs nanosheet can be evaluated. It is seen that when the simulated conduction band $E_C$ structures are tilted in the opposite ways, the evaluated electrical fields will point to opposite directions. To estimate the value of the Rashba coupling constant $\alpha$, we plot the experimentally extracted $\frac{1}{L_{SO}^2}$ as a function of $E^2$ in Fig. S5(e). It is seen that $\frac{1}{L_{SO}^2}$ shows a good linear dependence on $E^2$ at both $G \sim 7e^2/h$ and $G \sim 4e^2/h$. From the slopes of the linear dependences, the prefactors $r_R \sim 30.92$ e·nm$^2$ and $r_R \sim 40.42$ e·nm$^2$ can be extracted for the nanosheet at the device conductance values of $G \sim 7e^2/h$ and $G \sim 4e^2/h$. Figure S5(f) shows the absolute values of the Rashba coupling constant $|\alpha|$ as a function of $V_D$ for both conductance values. It is seen that $|\alpha|$ can be tuned from zero to a value of $\sim 0.38$ eVÅ at $G \sim 7e^2/h$, and from zero to a value of $\sim 0.42$ eVÅ at $G \sim 4e^2/h$. Thus, an efficient tuning of the Rashba coupling constant or the Rashba SOI strength in the InAs naonsheet with use of the dual gate has been demonstrated in our Device. Note that from the fits shown in Fig. S5(e), the values of $C_0$ could also be extracted out as $\sim 9$ μm$^{-2}$ at $G \sim 7e^2/h$ and $\sim 16$ μm$^{-2}$ at $G \sim 4e^2/h$. If we assume that the electric field-independent contribution to the spin procession originates dominantly from the Dresselhaus SOI in the InAs nanosheet, the Dresselhaus spin-orbit precession length can then be estimated as $L_{SO}^D \sim 0.33$ μm and $L_{SO}^D \sim 0.25$ μm in the nanosheet at $G \sim 7e^2/h$ and $G \sim 4e^2/h$, respectively.

**Section VI.** Electron transport measurement study of an additional dual-gate field-effect InAs nanosheet device (named as Device 2)

In this section, we present our measurement results obtained from Device 2, i.e., a similar dual-gate field-effect device made from another epitaxially grown InAs nanosheet and demonstrate that the device shows the same dual-gate voltage tunable magnetotransport characteristics as Device 1.

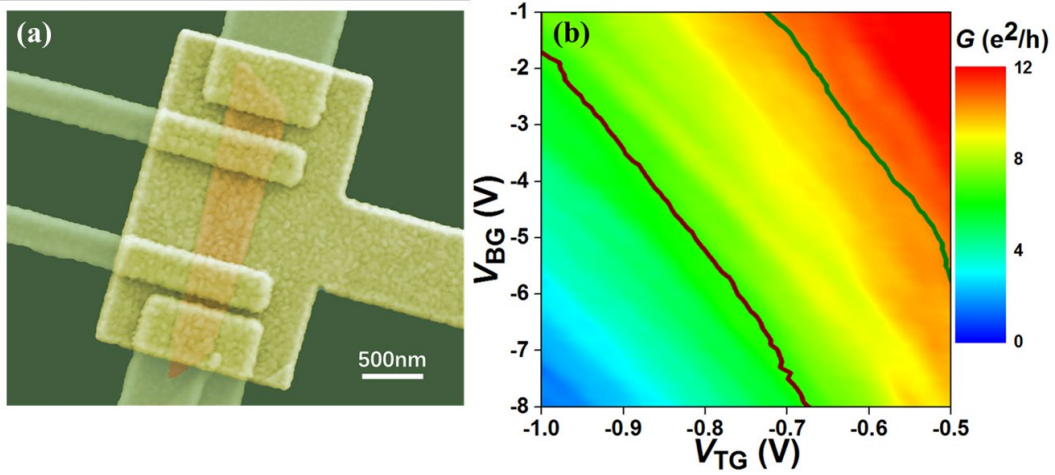

**FIG. S6.** (a) SEM image (top view, in false color) of Device 2, i.e., a dual-gate field-effect device with a similar device design as for Device 1 but made of another InAs nanosheet from the same MBE growth chip. (b) Conductance $G$ of Device 2 measured in a four-probe setup [as shown in Fig. 1(c) of the main article] as a function of $V_{BG}$ and $V_{TG}$ at $T = 2$ K. The green and brown solid lines represent the constant conductance contours of $G \sim 10.4e^2/h$ and $G \sim 6.5e^2/h$, respectively.

Figure S6(a) displays an SEM image (top view, in false color) of Device 2. The layer structure of the device is the same as it shown schematically in Fig. 1(d) of the main article. The InAs nanosheet employed in Device 2 is grown by molecular beam epitaxy (MBE) and is in fact taken from the same growth chip as the InAs nanosheet used in Device 1. The measurements are performed in the same circuit setup as for Device 1.

Figure S6(b) shows the measured dual-gate transfer characteristics of Device 2 at a temperature of $T = 2$ K. It is again seen that both $V_{BG}$ and $V_{TG}$ can effectively tune the conductance $G$ of the InAs nanosheet. The green and brown solid lines in Fig. S6(b) mark the constant conductance contours of $G \sim 10.4e^2/h$ and $G \sim 6.5e^2/h$, respectively.

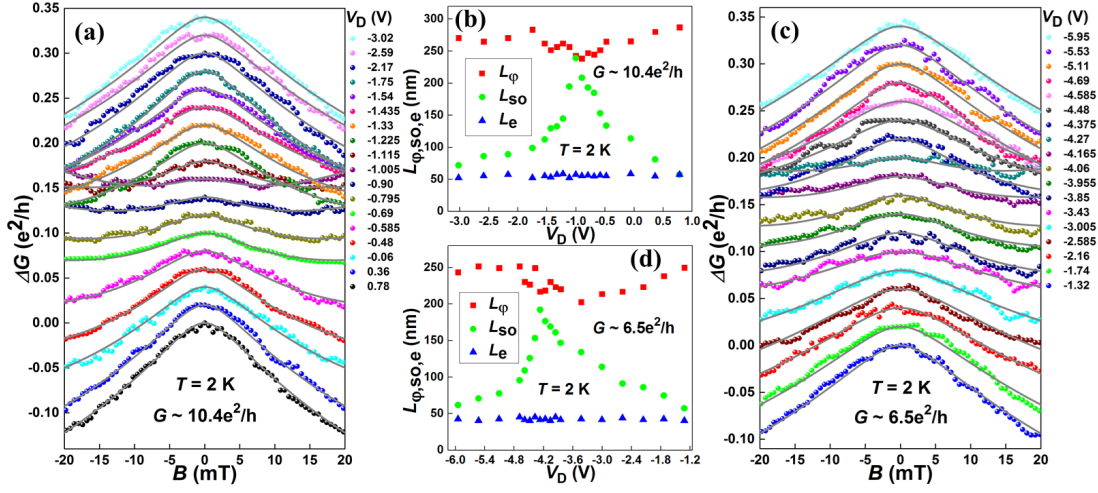

**FIG. S7.** (a) Low-field magnetoconductance $\Delta G$ measured for Device 2 at $T = 2$ K and at various dual-gate voltages $V_D$ taken along the conductance contour of $G \sim 10.4 e^2/h$. The bottom black data points display the measured $\Delta G$ data at $V_D = 0.78$ V and the $\Delta G$ data points measured at all other $V_D$ values are successively vertically offset by $0.02 e^2/h$ for clarity. The gray solid lines are the results of the fits of the measurements to the HLN theory [Eq. (1) in the main article]. (b) Characteristic transport lengths $L_\varphi$, $L_{SO}$, and $L_e$ as a function of $V_D$ extracted from the fits in (a). (c) The same as in (a) but for the dual-gate voltages $V_D$ taken along the conductance contour of $G \sim 6.5 e^2/h$. The bottom blue data points display the measured $\Delta G$ data at $V_D = -1.32$ V and the $\Delta G$ data measured at all other $V_D$ values are successively vertically offset by $0.02 e^2/h$ for clarity. Again, the gray solid lines are the results of the fits of the measurements to the HLN theory [Eq. (1) in the main article]. (d) Characteristic transport lengths $L_\varphi$, $L_{SO}$, and $L_e$ as a function of $V_D$ extracted from the fits in (c).

Figures S7(a) and S7(c) show the low-field magnetoconductance $\Delta G$ measured for Device 2 at $T = 2$ K and at different dual-gate voltages $V_D$ (defined as $V_D = V_{BG} - V_{TG}$) taken along the conductance contours of $G \sim 10.4 e^2/h$ and $G \sim 6.5 e^2/h$, respectively. It is seen that the measured $\Delta G$ data at every $V_D$ value display a peak-like structure near zero magnetic field, i.e., the WAL characteristics, indicating the presence of strong SOI in the InAs nanosheet. However, with decreasing $V_D$, the WAL peak feature is seen to become faint and then turn to get sharper and well-defined again in both the $G \sim 10.4 e^2/h$ and the $G \sim 6.5 e^2/h$ case. To extract characteristic transport lengths, the measured $\Delta G$ data are fitted to the HLN theory [Eq. (1) of the main article], see the solid lines in Figs. S7(a) and S7(c) for the results of the fits. Figures S7(b) and S7(d) show the extracted values of $L_\varphi$, $L_{SO}$, and $L_e$ as a function of $V_D$ from the fits shown in Figs. S7(a) and S7(c), respectively. It is clearly seen that both $L_\varphi$ and $L_e$ exhibit a weak dependence on dual-gate voltage $V_D$. However, the extracted $L_{SO}$ is modulated strongly by $V_D$ in the same way as in Device 1, namely, with decreasing $V_D$, $L_{SO}$ quickly increases to a peak value and then falls down quickly. This result demonstrates again that an SOI of the Rashba type is present in the InAs nanosheet in Device 2 and is greatly tunable with use of the dual gate. Note that when $L_{SO}$ is tuned to its peak value by $V_D$, the Rashba SOI should vanish, and the remnant SOI is dual-gate voltage independent and could dominantly be of the Dresselhaus type arising from bulk inversion asymmetry in the nanosheet. Note also that the phase coherence length $L_\varphi$ is comparably longer than the spin-orbit length $L_{SO}$ over the entire tuning range of dual-gate voltages $V_D$ in both the $G \sim 10.4e^2/h$ and the $G \sim 6.5e^2/h$ case. Thus, no crossover between $L_\varphi$ and $L_{SO}$ occurs with changing $V_D$. As a consequence, no WL characteristics should appear in the measured magnetoconductance of the InAs nanosheet, which is fully consistent with the measurement results shown in Figs. S7(a) and S7(c).

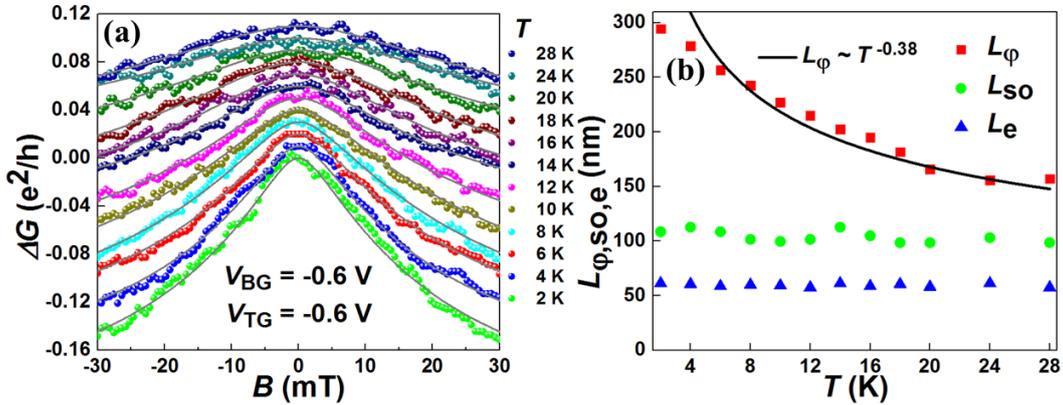

**FIG. S8.** (a) Low-field magnetoconductance $\Delta G$ measured for Device 2 at different temperatures $T$ at fixed gate voltages $V_{BG} = -0.6$ V and $V_{TG} = -0.6$ V. The bottom green data points display the measured $\Delta G$ data at $T = 2$ K and the $\Delta G$ points measured at all other temperatures are successively vertically offset by $0.01e^2/h$ for clarity. The gray solid lines are the results of the fits of the measurements to the HLN theory [Eq. (1) in the main article]. (b) Characteristic transport lengths $L_\varphi$, $L_{SO}$, and $L_e$ as a function of $T$ extracted from the fits in (a). The black solid line is a power-law fit of the extracted $L_\varphi$, showing a $L_\varphi \sim T^{-0.38}$ temperature dependence.

Figure S8(a) shows the temperature-dependent measurements of the low-field magnetoconductance $\Delta G$ for Device 2 at fixed gate voltages $V_{BG} = -0.6$ V and $V_{TG} = -0.6$ V. Here, it is seen that all the measured $\Delta G$ data at temperatures ranging from 2 to 28 K show a peak-like structure near zero magnetic field, i.e., the WAL characteristics. Nevertheless, the WAL peak is sharp at $T = 2$ K and is gradually broadened with increasing $T$. Figure S8(b) shows the extracted $L_\varphi$, $L_{SO}$ and $L_e$ as a function of $T$ from the fits of the measured data to the HLN theory [Eq. (1) in the main article]. It is seen that both $L_{SO}$ and $L_e$ show a weak temperature dependence, while $L_\varphi$ decreases rapidly with increasing temperature. It is also found that the extracted $L_\varphi$ in the InAs nanosheet can be fitted by a power-law temperature dependence of $T^{-0.38}$ [see the solid line in Fig. S8(b)]. This is the same result as found for Device 1 in the main article and again implies that the dephasing in the InAs nanosheet arises dominantly from electron-electron interactions with small energy transfers (the Nyquist scattering mechanism).